\documentclass[aps,prl,twocolumn,showpacs,superscriptaddress,bibliography]{revtex4-1}
\usepackage{graphicx}    
\usepackage{dcolumn}    
\usepackage{bm}             
\usepackage{amssymb}   
\usepackage{amsmath}    
\usepackage{subfigure}    
\usepackage{braket}
\usepackage{mathrsfs}
\usepackage[export]{adjustbox}
\usepackage[toc,page]{appendix}
\begin{document}

\title{Motional squeezing for trapped ion transport and separation}

\author{R. T. Sutherland}
\email{robert.sutherland@utsa.edu}
\affiliation{Department of Electrical and Computer Engineering, University of Texas at San Antonio, San Antonio, Texas 78249, USA}
\author{S. C. Burd}
\affiliation{Time and Frequency Division, National Institute of Standards and Technology, 325 Broadway, Boulder, Colorado 80305, USA}
\affiliation{Department of Physics, University of Colorado, Boulder, Colorado 80309, USA}
\author{D. H. Slichter}
\affiliation{Time and Frequency Division, National Institute of Standards and Technology, 325 Broadway, Boulder, Colorado 80305, USA}
\author{S. B. Libby}
\affiliation{Physics Division, Physical and Life Sciences, Lawrence Livermore National Laboratory, Livermore, California 94550, USA}
\author{D. Leibfried}
\affiliation{Time and Frequency Division, National Institute of Standards and Technology, 325 Broadway, Boulder, Colorado 80305, USA}
\date{\today}

\begin{abstract}
Transport, separation, and merging of trapped ion crystals are essential operations for most large-scale quantum computing architectures. In this work, we develop a theoretical framework that describes the dynamics of ions in time-varying potentials with a motional squeeze operator, followed by a motional displacement operator. Using this framework, we develop a new, general protocol for trapped ion transport, separation, and merging. We show that motional squeezing can prepare an ion wave packet to enable transfer from the ground state of one trapping potential to another. The framework and protocol are applicable if the potential is harmonic over the extent of the ion wave packets at all times. As illustrations, we discuss two specific operations: changing the strength of the confining potential for a single ion, and separating same-species ions with their mutual Coulomb force. Both of these operations are, ideally, free of residual motional excitation.
\end{abstract}
\maketitle

A suitable platform for quantum information processing must enable the precise control of many-body quantum states \cite{nielsen_2010, ladd_2010}. Trapped ions are promising in this regard due to their long coherence times, high-fidelity gate operations, and potential for all-to-all connectivity between qubits \cite{cirac_1995, monroe_1995,wineland_1998,haffner_2008, blatt_2008,harty_2014,ballance_2016,gaebler_2016,wang_2017}. One way to address the challenge of scaling to larger trapped-ion systems is the so-called quantum charge-coupled device (QCCD). In the QCCD architecture, ions are shuttled between different trap `zones' that can have designated functions such as gate operations, memory, or readout \cite{wineland_1998,kielpinski_2002,pino_2020}. To be as efficient as possible, separation and transport of same and mixed ion species should be fast and minimize residual motional excitation. While theoretical work has explored various shuttling protocols \cite{torrontegui_2011, lau_2011, kaufmann_2014,palmero_2014,palmero_2015,ruster_2014, lau_2014}, only single-ion and same-species ion transport have been demonstrated on timescales comparable to the ion's motional period; experimentalists have performed other operations, but only adiabatically with respect to the motional period \cite{blakestad_2009,bowler_2012,ruster_2014}. 

\begin{figure}[tb]
\includegraphics[width=0.49\textwidth]{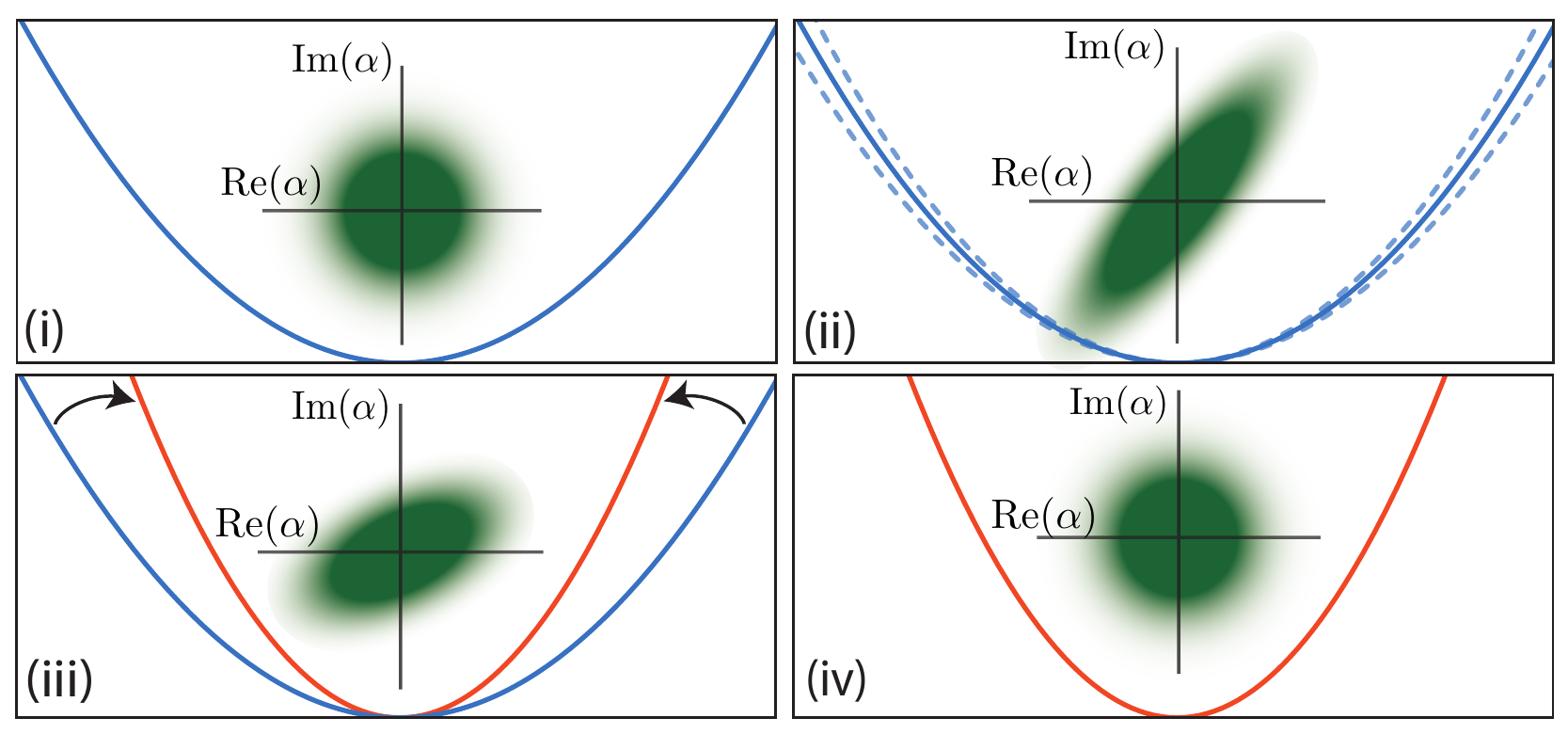}
\centering
\caption{We illustrate a change of the potential of a harmonic trap, showing the motion's Wigner distribution $W(\alpha)$ \cite{agarwal_2012} for each step (not to scale). We (i) initialize the motion to the ground state of a potential well, (ii) squeeze the motion (shown as parametric modulation of the potential), and (iii) increase the confinement strength. Finally, (iv) the motion finishes in the ground state of the final potential.}
\label{fig:pretty}
\end{figure}

In this work, we develop a new theoretical framework to analyze the motional states of ions in a linear trap with time-varying potentials.  Specifically, we consider the case of ions starting and ending in the ground states of a set of harmonic wells with frequencies and equilibrium positions $\{\omega_{j}(0),c_{j}(0)\}$ to $\{\omega_{j}(t_{f}),c_{j}(t_{f})\}$ over duration $t_{f}$, where $j$ indicates the motional mode (see Fig.~\ref{fig:pretty}). This framework can be applied if, at all times, the effective potential can be approximated as quadratic over the spatial extent of the ion motional wave packets. Under this condition, the wave packets remain Gaussian and follow classical trajectories \cite{ehrenfest_1927,heller_1975, huber_1987, garraway_2000}. This fact allows us to account for the `classical' position and momentum of the ions with a frame transformation, given by a displacement operator. In this `classical frame of reference', the Hamiltonian can be cast in a basis that is represented by generators of the SU(1,1) Lie algebra \cite{perelomov_1972,wodkiewicz_1985,gerry_1985,yurke_1986, wu_2006}, allowing us to reduce the remaining dynamics to Euler rotations \cite{holman_1966,wodkiewicz_1985,woit_2017}. Once the angles of these rotations are determined, the effect of the entire operation is equivalent to a squeezing operation, followed by a coherent displacement; this operator parameterization of Gaussian trajectories has been used in other contexts \cite{garraway_2000, lau_2012}, but its use in QCCD operations has not been explored. Further, we can add an additional squeezing operation per mode, before or after the main dynamical operation, so that the system finishes in the ground states of a set of final trapping potentials with arbitrary frequencies $\omega_{j}(t_{f})$. Taken together, one can transport ions from the ground states of one set of wells to the ground states of another, with no motional excitation. We present two applications of this technique: changing the frequency of an ion's motion, and separating two same-species ions. Using experimentally realistic parameters, we show that the latter could potentially be conducted more than an order of magnitude faster than in previous demonstrations. 

We consider $N$ ions in a linear trap, described by `lab-frame' motional wave function $\ket{\psi(t)}$ and Hamiltonian:
\begin{equation}\label{eq:ham_orig}
    \hat{H}_{l}(t) = \sum_{j}\frac{\hat{p}^{2}_{j}}{2m_{j}} + V_{j}(\hat{x}_{j},t),
\end{equation}
where $\hat{p}_{j}$ is momentum, $\hat{x}_{j}$ is position, $m_{j}$ is mass, and $t$ is time. The index $j$ can here indicate the coordinates of an individual ion or a collective mode, depending on the problem. We assume negligible coupling between different degrees of freedom; this approximation means that for systems with more than one ion we can factor $\ket{\psi(t)}$. Therefore, each degree of freedom $j$ can be considered separately, and we drop this subscript unless stated otherwise. We now transform $\hat{H}_{l}(t)$ into a frame of reference that is centered around the `classical' position $c(t)$ and momentum $m\dot{c}(t)$. This unitary transformation is represented by the displacement operator $\hat{D}(t) \equiv \exp\{(i/\hbar)[m\dot{c}(t)\hat{x} - c(t)\hat{p}]\}$. This gives  \cite{wolf_1995} $\hat{D}^{\dagger}(t)\hat{x}\hat{D}(t) = \hat{x} + c(t)$ and $\hat{D}^{\dagger}(t)\hat{p}\hat{D}(t) = \hat{p} + m\dot{c}(t)$. We consider the transformed wave function $\ket{\phi(t)}=\hat{D}^{\dagger}(t)\ket{\psi(t)}$ and Hamiltonian $\hat{H}_{t}(t)=\hat{D}^{\dagger}(t)\hat{H}_{l}(t)\hat{D}(t) + i\hbar\dot{\hat{D}}^{\dagger}(t)\hat{D}(t)$, which gives \cite{supplemental}:
\begin{eqnarray}\label{eq:classical_frame}
    \hat{H}_{t}(t) &= & \frac{\hat{p}^{2}}{2m} + V(\hat{x},t) - \hat{x}\frac{\partial V}{\partial \hat{x}}|_{\hat{x}=0}.
\end{eqnarray}
If we assume the potential $V(\hat{x},t)$ is quadratic around $c(t)$, we can write the above equation as:
\begin{eqnarray}\label{eq:trans_h}
    \hat{H}_{t}(t) \simeq \frac{\hat{p}^{2}}{2m} + \frac{1}{2}m\omega^{2}(t)\hat{x}^{2}.
\end{eqnarray}
Here, we have set $m\omega^{2}(t) \equiv \partial^{2}V/\partial\hat{x}^{2}|_{\hat{x}=0}$. Notice that Eq.~(\ref{eq:trans_h}) does not have a $\propto \hat{x}$ force term, as its effect is now encompassed by $\hat{D}(t)$. This gives a free harmonic oscillator with a time-dependent potential centered at $\braket{\hat{x}}=0$. 

We are interested in modes that begin in a $\omega(0) \equiv\omega_{0}$ potential well. We set $\omega(t)\equiv\{1+\gamma(t)\}\omega_{0}$, where $\gamma(t)$ is a dimensionless time-dependent function. We rewrite Eq.~(\ref{eq:trans_h}) in terms of ladder operators $\hat{a}(\hat{a}^{\dagger})$ acting on the mode defined by $\omega(0)$. Doing this, and ignoring global phases, gives \cite{supplemental}: 
\begin{eqnarray}\label{eq:main_ham}
    \hat{H}_{t}(t) &=& \hbar\omega_{0}\Big\{\hat{a}^{\dagger}\hat{a} +  \frac{\gamma(t)}{2}\Big[1 \nonumber  + \frac{\gamma(t)}{2}\Big]\Big[\hat{a}^{\dagger} + \hat{a}\Big]^{2}\Big\} \nonumber \\
    &= & 2\hbar\omega_{0}\Big\{[1+\alpha(t)]\hat{K}_{3} + \alpha(t)\hat{K}_{1}\Big\},
\end{eqnarray}
where we have substituted $\hat{x} \equiv (\hbar/2m\omega_{0})^{1/2}(\hat{a}^{\dagger}+\hat{a})$, $\hat{p} \equiv i(\hbar m\omega_{0}/2)^{1/2}(\hat{a}^{\dagger}-\hat{a})$, and $\alpha(t) \equiv \gamma(t)\{1 + \frac{1}{2}\gamma(t)\}$. Importantly, Eq.~(\ref{eq:main_ham}) introduces the generators of the SU(1,1) Lie algebra \cite{perelomov_1972,wodkiewicz_1985,gerry_1989_2}:
 \begin{eqnarray}
    \hat{K}_{1} &\equiv & \frac{1}{4}\{\hat{a}^{\dagger 2} + \hat{a}^{2}\}, 
    \hat{K}_{2} \equiv \frac{1}{4i}\{\hat{a}^{\dagger 2} - \hat{a}^{2}\},
    \hat{K}_{3} \equiv \frac{1}{2}\{\hat{a}^{\dagger}\hat{a} + \frac{1}{2}\}. \nonumber \\
\end{eqnarray}
Because Eq.~(\ref{eq:main_ham}) depends only on these generators, we may represent the propagator associated with $\hat{H}_{t}(t)$ as $\hat{U}_{s}$ with three Euler rotations in SU(1,1) space \cite{wodkiewicz_1985,holman_1966}:
\begin{equation}
    \hat{U}_{s} = e^{i\theta_{a}\hat{K}_{3}}e^{2ir_{s}\hat{K}_{2}}e^{i\theta_{b}\hat{K}_{3}},
\end{equation}
where we have defined the angle $r_{s}$ such that its value corresponds to the squeeze parameter \cite{agarwal_2012}. We designate the position and velocity coordinates of the final potential minimum (in the lab-frame) as $c_{f}$ and $\dot{c}_{f}$, respectively, distinct from the ion coordinates $c(t_{f})$ and $\dot{c}(t_{f})$, to encompass residual displacements in our framework; when $c(t_{f})={c}_{f}$, the packet is centered at the final potential minimum, and, when $\dot{c}(t_{f})=\dot{c}_{f}$, the ion is stationary with respect to it. Transforming into the reference frame centered at, and stationary with respect to, the minimum of the final potential after duration $t_{f}$ gives $\ket{\phi_{f}(t_{f})}=\hat{D}_{f}^{\prime}\hat{U}_{s}\ket{\phi(0)}$, where the final frame change is represented by the displacement operator $\hat{D}_{f}^{\prime}\equiv\exp\{(i/\hbar)[m\{\dot{c}(t_{f})-\dot{c}_{f}\}\hat{x} - \{c(t_{f})-c_{f}\}\hat{p}]\}$ \cite{supplemental}. Under the approximation that $V(\hat{x},t)$ is quadratic over the spatial the extent of $\ket{\phi(t)}~\forall t$, we can thus represent ion motional dynamics with a squeeze, followed by a displacement, operator. While there is a broad set of transport, separation, and mode frequency change operations this framework could analyze, for this work we consider only the subset of operations where $c(t_{f})=c_{f}$ and $\dot{c}(t_{f})=\dot{c}_{f}=0$, from which it follows that $\hat{D}_{f}^{\prime}=\hat{I}$; this framework could, however, straightforwardly describe protocols where ions are caught by moving potentials ($\dot{c}_{f}\neq 0$), or are displaced at $t_{f}$ ($\hat{D}_{f}^{\prime}\neq \hat{I}$). 

As an example, we can study the case of taking the system from the ground state of a well with frequency and coordinate $\{\omega(0), c(0)\}$ to the ground state of $\{\omega(t_{f}), c_{f}\}$. Finding experimentally realistic functions that efficiently take the system from $c(0)$ to $c_{f}$ while \textit{simultaneously} taking the system from the ground state of $\omega(0)$ to the ground state of $\omega(t_{f})$ is a difficult task in general; we can, however, guarantee the latter requirement by introducing an additional step, either before or after the transport, separation, or frequency change operation, that squeezes the motion so it ends in the ground state of the $\omega(t_{f})$ potential well. This works so long as $c(t_{f}) = c_{f}$ and $\dot{c}(t_{f}) = \dot{c}_{f}$. The operator that describes changing the mode from the ground state of $\omega(0) = \omega_{0}$ to that of $\omega(t_{f}) = \{1 + \gamma(t_{f})\}\omega_{0}$ is \cite{supplemental}:
\begin{equation}\label{eq:change_freq_op}
   \hat{U}_{c} = e^{2ir_{c}\hat{K}_{2}},~ r_{c} \equiv -\frac{1}{2}\ln\Big\{1 +\gamma(t_{f})\Big\}.
\end{equation}
We want to find a squeezing operation $\hat{U}_{p}$ that, when applied before or after the main dynamical operation operation, gives the desired mode frequency change. In equation form, this is $\hat{U}_{s}\hat{U}_{p} = \hat{U}_{c}$ or $\hat{U}_{p}\hat{U}_{s}=\hat{U}_{c}$ if squeezing is applied before or after the main operation, respectively. We can split $\hat{U}_{p}$ into Euler rotations:
\begin{eqnarray}
    \hat{U}_{p} &=& e^{i\theta_{m}\hat{K}_{3}}e^{2ir_{p}\hat{K}_{2}}e^{i\theta_{n}\hat{K}_{3}}.
\end{eqnarray}
Concentrating on squeezing before the main operation, we can take $\hat{U}_{p}$ to act on $\ket{\phi(0)}=\ket{0}$. Up to a global phase, this gives:
\begin{eqnarray}\label{eq:prep_squeeze}
    \hat{U}_{p}\ket{0} &=& \exp\Big(\frac{r_{p}}{2}\Big\{\hat{a}^{\dagger 2}e^{i\theta_{m}}-\hat{a}^{2}e^{-i\theta_{m}}\Big\}\Big)\ket{0}.
\end{eqnarray}

Squeezing can be induced with parametric modulation or a diabatic change of the trapping potential using applied voltages \cite{heinzen_1990,burd_2019, heinzen_1990,wittemer_2019}, or with lasers \cite{cirac_1993, meekhof_1996, kienzler_2015, dupays_2020}. The focus of this work is not how squeezing is generated, and, importantly, the validity of the above scheme does not depend on the timescales of the squeezing generation. We, therefore, assume a general squeezing Hamiltonian $\hat{H}_{I}$ considered in the interaction picture of $\hbar\omega_{0}\hat{a}^{\dagger}\hat{a}$:
\begin{equation}
    \hat{H}_{I} = i\hbar\frac{g}{2}\Big\{\hat{a}^{\dagger 2}e^{i\theta_{I}} - \hat{a}^{2}e^{-i\theta_{I}} \Big\},
\end{equation}
which describes the squeezing induced by frequency modulation of the trap frequency at $2\omega_{0}$ (see Fig.~\ref{fig:pretty}). After a duration $t_{p}$ the propagator representing the lab-frame action of $\hat{H}_{I}$ is, up to a phase,
\begin{eqnarray}
    \hat{U}_{I}\ket{0} &=& \exp\Big(\frac{g t_{p}}{2}\Big\{\hat{a}^{\dagger 2}e^{i[\theta_{I}-2\omega t_{p}]} - \hat{a}^{2}e^{-i[\theta_{I} - 2\omega t_{p}]}\Big\} \Big)\ket{0}, \nonumber \\
\end{eqnarray}
which is equivalent to $\hat{U}_{p}$ when $g t_{p} = r_{p}$ and $\theta_{I} = 2\omega_{0} t_{p} + \theta_{m}$; this means that squeezing the ground state of $\{\omega(0),c(0)\}$ can prepare it to end in that of $\{\omega(t_{f}),c_{f}\}$ after a change of the external potential. We provide two examples of how this technique may be used below. 

\begin{figure}
\includegraphics[width=0.49\textwidth]{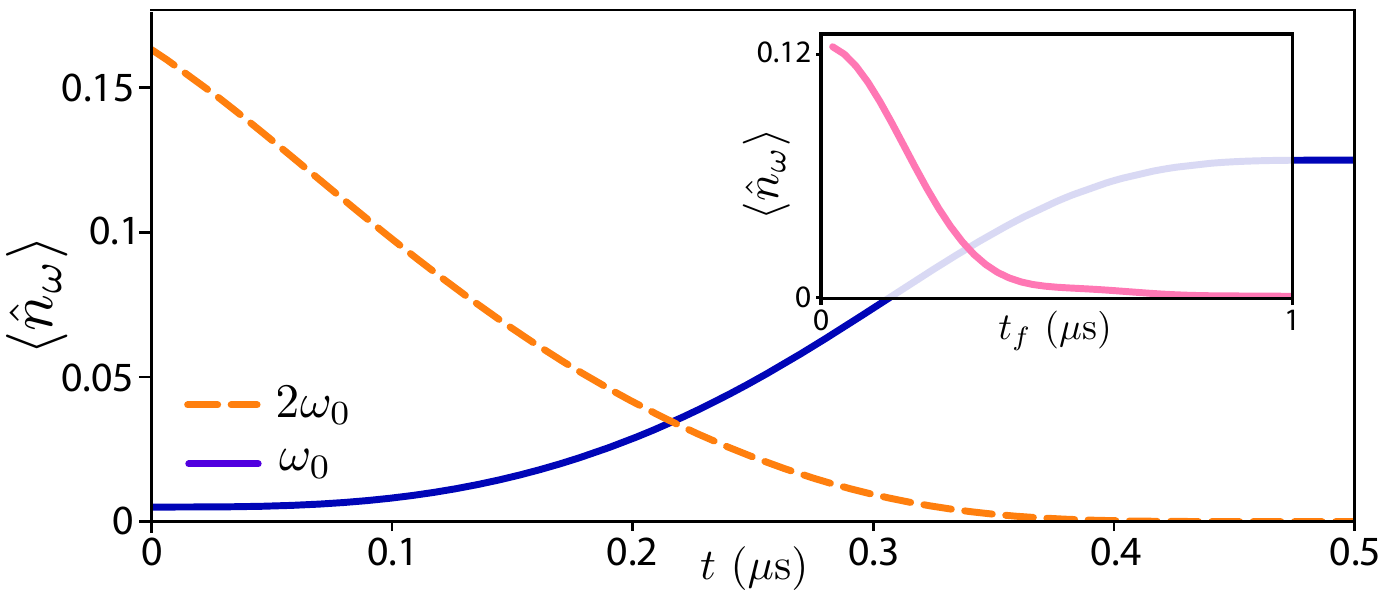}
\centering
    \caption{Changing the frequency from $\omega_{0}$ to $2\omega_{0}$ ($\omega_{0}/2\pi = 1 ~\mathrm{MHz}$) of a harmonic trap after a squeezing operation. We show the phonon number $\braket{\hat{n}_{\omega}}$ in the eigenbases of the original mode at $\omega_{0}$ (solid blue) and the final mode at $2\omega_{0}$ (dashed orange). The inset shows $\braket{\hat{n}_{\omega}}$ versus $t_{f}$ in the eigenbasis of the final well when no squeezing is performed; if squeezing were performed $\braket{\hat{n}_{\omega}}$ would remain zero for all $t_{f}$.}
\label{fig:change_freq}
\end{figure}

We first discuss how to use squeezing to change the motional frequency of an ion in a harmonic potential without residual motional excitation. We choose this as an initial example because it is a simple illustration of how squeezing can not only transform a wave packet from the ground state of one well to another, but also account for the change of the external potential, where $\hat{H}_{l}(0)\rightarrow\hat{H}_{l}(t_{f})$ over a finite duration. Figure~\ref{fig:pretty} pictorially illustrates this scheme. In this example, we increase the frequency of a mode, squeezed beforehand according to $\hat{U}_{p}$, following the function: $\gamma(t) = \sin^{2}(\pi t/2t_{f})$ in which the trap frequency is doubled from $\omega_{0}$ to $2\omega_{0}$ over a time period $t_{f}$. The functional form of $\gamma(t)$ can be chosen arbitrarily; different functional choices would not qualitatively affect the results, so long as boundary conditions remain satisfied. Here, the classical position of the particle remains at rest, trivially meeting the requirement that $c(t_{f}) = c_{f}$ and $\dot{c}(t_{f}) = \dot{c}_{f}$. Figure~\ref{fig:change_freq} shows the phonons $\braket{\hat{n}_{\omega}}$ (as defined by the ladder operators of the initial $\omega_{0}/2\pi = 1 ~\mathrm{MHz}$ and final $\omega(t_{f})/2\pi=2 ~\mathrm{MHz}$ trapping frequencies \cite{supplemental}) versus time for a ramp time of $t_{f} = 0.5~\mu\mathrm{s}$. Here we see that the motion is not in the ground state of either basis after the initial squeezing operation; after doubling the frequency, however, the final well is. 

The inset of Fig.~\ref{fig:change_freq} shows the residual phonons in the $2\omega_{0}$ mode versus $t_{f}$ \textit{without} a squeezing step $\hat{U}_{p}$. It is interesting to note that, when $t_{f}\rightarrow 0$, the number of residual phonons approaches that of the $\omega_{0}$ mode in the main figure. This is because, in this regime, $\hat{U}_{s}\rightarrow \hat{I}$; this means the $\hat{U}_{p}$ operation, that \textit{would} transform the wave function from the ground state of $\omega_{0}$ to that of $2\omega_{0}$, converges to $\hat{U}_{c}$. Since experimental techniques for squeezing typically do not operate in time frames shorter than $2\pi/\omega_{0}$, the inset indicates that our squeezing scheme is unlikely to be more time-efficient than just adiabatically changing the potential, so this offers only a simple example of the protocol. Ion separation, on the other hand, could potentially be expedited with squeezing.

\begin{figure}[tb]
\includegraphics[width=0.49\textwidth]{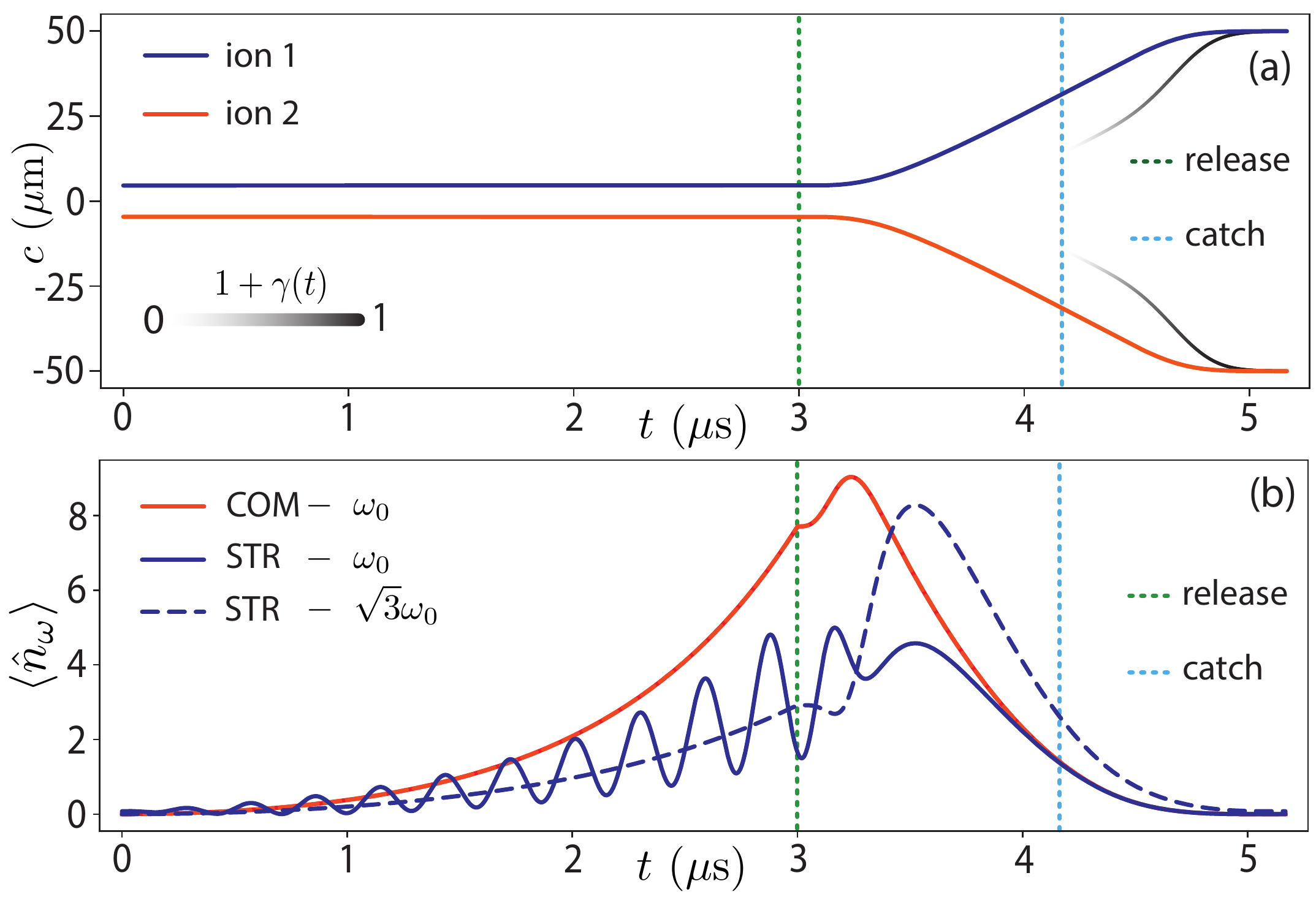}
\centering
\caption{Illustration of same-species separation using squeezing. Both figures are for the same run, using parametric modulation with amplitudes of $g_{c}/2\pi \simeq 92.6~\mathrm{kHz}$ and $g_{s}/2\pi\simeq 69.2 ~\mathrm{kHz}$ acting simultaneously on the center-of-mass (COM) and stretch (STR) modes, respectively. Here $\omega_{0}/2\pi = 1~\mathrm{MHz}$. This example protocol is comprised of a $t_{p} = 3~\mu\mathrm{s}$ parametric modulation step, followed by $t_{s,1} = 0.5~\mu\mathrm{s}$ of ramping down the original confining potential to $\omega(t_{p}+t_{s,1})=0$, $t_{s,2} \simeq 0.67~\mu\mathrm{s}$ of only Coulomb repulsion, followed by $t_{s,3} = 1~\mu\mathrm{s}$ of ramping up the `catching' potential. This gives a total separation of $c_{1}(t_{f}) - c_{2}(t_{f})\simeq 100~\mu\mathrm{m}$ over $t_{f}\simeq 5.17~\mu\mathrm{s}$. (a) Shows the respective position of ions $1$ and $2$ versus time $t$ in microseconds. The green and blue dotted lines mark the beginnings of the release and catch steps, respectively. The black lines show the position of the catching potential well versus $t$ and the lines' shade is proportional to the fraction of its final strength, $1+\gamma(t)$. (b) Shows the phonons in the COM mode, and both of the relevant STR modes versus time. At the end of the trajectory, both ions are in the ground state of their spatial separated $\omega(t_{f}) = \omega_{0}$ potential wells, while the $\sqrt{3}\omega_{0}$ mode is not.
}
\label{fig:sep}
\end{figure}

We now discuss a protocol that uses motional squeezing to diabatically separate two same-species ions. We begin with two ions, here taken to be $^{9}\mathrm{Be}^{+}$, in an initial potential well with frequency $\omega_{0}$ at the equilibrium positions of ions $1$ and $2$ at $\sqrt[3]{ke^{2}/(4m\omega_{0}^{2})}$ and $-\sqrt[3]{ke^{2}/(4m\omega_{0}^{2})}$, respectively, where $k$ is the electrostatic constant and $e$ is charge. We can then take the usual coordinate system $\hat{x}_{c}=\frac{1}{2}(\hat{x}_{1}+\hat{x}_{2})$ and $\hat{p}_{c} = \hat{p}_{1} + \hat{p}_{2}$ for the center-of-mass (COM) mode, $\hat{x}_{s}=\frac{1}{2}(\hat{x}_{1}-\hat{x}_{2})$ and $\hat{p}_{s}= \hat{p}_{1} - \hat{p}_{2}$ for the stretch (STR) mode, each with effective mass $M=2m$. This allows us to analyze the system in an uncoupled basis, $\hat{H}_{t}(t) = \hat{H}_{t,c}(t) + \hat{H}_{t,s}(t)$. Dropping terms $\propto \hat{I}$, we get:
\begin{eqnarray}\label{eq:dif_ham}
    \hat{H}_{t,c}(t) &=& \frac{\hat{p}_{c}^{2}}{2M} + \frac{1}{2}M\omega_{0}^{2}\Big\{1 + \gamma(t) \Big\}^{2}\hat{x}_{c}^{2} \label{eq:com_ham}
\end{eqnarray}  
\begin{eqnarray}
    \hat{H}_{t,s}(t) &=& \frac{\hat{p}_{s}^{2}}{2M} + \frac{1}{2}M\omega_{0}^{2}\Big\{1 + \gamma_{s}(t) \Big\}^{2}\hat{x}_{s}^{2}.
\end{eqnarray}
We have Taylor-expanded the Coulomb potential up to $\hat{x}_{s}^{2}$, encompassing this term's dynamics in \cite{supplemental}:
\begin{equation}
    \gamma_{s}(t) \equiv \Big\{[1 + \gamma(t)]^{2} + \frac{2c_{s}^{3}(0)}{c_{s}^{3}(t)} \Big\}^{1/2} - 1,
\end{equation}
in order to cast $\hat{H}_{t,c}(t)$ and $\hat{H}_{t,s}(t)$ in the same form.  Initially, the STR mode frequency is $\omega_{s}(0) = \sqrt{3}\omega_{0}$, so we measure the phonon number $\braket{\hat{n}_{\omega}}$ as defined by these ladder operators, while measuring the COM modes in terms of $\omega_{c}(0) = \omega_{0}$ operators. When separating into different trap zones, however, the $\propto c_{s}^{-3}(t)$ Coulomb term in $\gamma_{s}(t)$ becomes negligibly small, giving $\omega_{s}(t_{f}) \simeq \omega_{0}$; this makes $\braket{\hat{n}_{\omega}} = 0$, defined by $\omega_{s}(t_{f}) \simeq \omega_{c}(t_{f}) = \omega_{0}$ ladder operators, the target for both modes.

The separation of the classical trajectory from $\mathrm{SU}(1,1)$ dynamics allows us to isolate each when designing a protocol. Therefore, we discuss individual positions for the former and modes for the latter. We first choose a protocol that separates the particles from $\{c_{1}(0),c_{2}(0) \}$ to a desired $\{c_{f,1}(t_{f}),c_{f,2}(t_{f}) \}$, such that the particles finish in equilibrium. After the parametric modulation sequence lasting $t_{p}$, we release the ions from confinement, ramping the potential to zero over a duration $t_{s,1}$ according to $\gamma(t) = -\sin^{2}(\pi[t-t_{p}]/2t_{s,1})$. Subsequently, for a duration $t_{s,2}$ we leave the particles unconfined ($\gamma(t) = -1$), after which we apply separate `catching' potentials over duration $t_{s,3}$ according to $\gamma(t) = -\cos^{2}(\pi[t - \{t_{p} + t_{s,1} + t_{s,2}\}]/2t_{s,3})$; $\gamma(t) = 0$ everywhere else. We set the centers of the `catching' potentials to be $c_{f,j}(t) = c_{j}(t) - \eta\dot{c}_{j}(t)$, where $\eta$ is a constant with dimensions of time. This ensures that $c_{j}(t_{f}) \simeq c_{f,j}(t_{f})$ and $\dot{c}_{j}(t_{f}) \simeq \dot{c}_{f,j}(t_{f}) \simeq \ddot{c}_{j}(t_{f}) \simeq \ddot{c}_{f,j}(t_{f}) \simeq 0$, whereby $\hat{D}^{\prime}_{f,j}\rightarrow\hat{I}$. Figure~\ref{fig:sep}a shows the classical trajectories of two ions being separated by $c_{1}(t_{f}) - c_{2}(t_{f}) \simeq 100~\mu\mathrm{m}$ in $t_{s,1} + t_{s,2} + t_{s,3} \simeq 2.17~\mu\mathrm{s}$, not including the squeezing period. Here we have set $\omega_{0}/2\pi = 1~\mathrm{MHz}$, $t_{p} = 3~\mu\mathrm{s}$, $t_{s,1} = 0.5~\mu\mathrm{s}$, $t_{s,2} \simeq 0.67~\mu\mathrm{s}$, $t_{s,3} = 1~\mu\mathrm{s}$, and $\eta = 0.5~\mu\mathrm{s}$\textemdash the values of $g$ for both modes are determined after the values of $r_{p}$ are calculated (discussed below). The ions remain at $c_{j}(0)$ during the squeezing stage, then quickly separate when their initial confinement is dropped, coming to rest after their catching potentials reach their full value at $t_{f}$.

The design of a scheme where both ions come to rest at their respective potential minimums requires many adjustable parameters. The squeezing needed to prepare each mode for separation, however, is virtually identical to that in our discussion of changing the mode frequency, only with different $\gamma(t)$. Here, we squeeze both modes simultaneously for a fixed $t_{p}$, making the required values of $g_{c}$ and $g_{s}$ different. For the values of $r_{p,c}$ and $r_{p,s}$ in the example shown in Fig.~\ref{fig:sep}, we find that $g_{c}/2\pi \simeq 93~\mathrm{kHz}$ and $g_{s}/2\pi\simeq 69~\mathrm{kHz}$. These values of $g_{s,c}$ were chosen to correspond to current state-of-art experiments \cite{burd_2019, burd_2020}, but are not necessary for this scheme to work; the use of stronger or weaker squeezing $g_{s,c}$ would simply cause $t_p$ to scale as $1/g_{s,c}$. For this calculation, $\omega_{c}(0) = \omega_{c}(t_{f})$, but $\omega_{s}(0)\neq \omega_{s}(t_{f})$. To finish in the ground state of $\omega_{s}(t_{f}) = \omega_{0}$, we incorporate this change of frequency into $\hat{U}_{p,s}$, such that the wave packet changes from the ground state of $\sqrt{3}\omega_{0}$ to that of $\omega_{0}$. In Fig.~\ref{fig:sep}b, we show $\braket{\hat{n}_{\omega}}$ for the modes defined by $\omega_{c}(0) = \omega_{c}(t_{f})$, $\omega_{s}(0)$, and $\omega_{s}(t_{f})$ versus time for the same separation shown in Fig.~\ref{fig:sep}a. This shows that both modes end in the ground state of their final potentials. For this protocol, we see the largest squeezing parameter is $r_{p,c}\simeq 1.8$, which is experimentally feasible \cite{burd_2019}.

In conclusion, this work presents a new, general method for analyzing the behavior of ions in time-varying potentials, and for designing improved ion transport, separation, and merging protocols by using motional squeezing. First, we show that when the Hamiltonian of an ion or ions in a time-varying potential takes the form of Eq.~(\ref{eq:main_ham}), after a frame transformation that accounts for the classical trajectory of each ion, the remaining dynamics of the system can be described by three Euler rotations in $\mathrm{SU}(1,1)$ space. When acting on the ground state of a motional mode, we show that one can use a single squeezing operation per mode such that the wave packet finishes its trajectory in the ground state of the final potential. It is important to note that the frequency change and separation protocols shown above represent specific examples of a wide range of feasible transport, separation, or merging schemes described by Eq.~(\ref{eq:main_ham}). It is reasonable to expect that variants of these two examples would perhaps better suit a particular experimental setup, or that other types of transport, separation, or mode frequency change operations may be catalyzed by this concept. This work could, therefore, open many new options for designing schemes that use motional squeezing in QCCD operations.

We would like to thank F. Robicheaux, D. J. Wineland, R. Srinivas, and S. L. Todaro for helpful discussions, and P. Hou and A. Kwiatkowski for comments on the manuscript. S. C. B. is an Associate in the Professional Research Experience Program (PREP) operated jointly by  NIST and the University of Colorado Boulder under award 70NANB18H006 from the U.S. Department of Commerce, National Institute of Standards and Technology. This work was supported by the NIST Quantum Information Program. Part of this work was performed under the auspices of the U.S. Department of Energy by Lawrence Livermore National Laboratory under Contract  DE-AC52-07NA27344.

\bibliography{biblio}

\end{document}


\title{Supplemental Material: Motional squeezing for trapped ion transport and separation}

\author{R. T. Sutherland}
\email{robert.sutherland@utsa.edu}
\affiliation{Department of Electrical and Computer Engineering, University of Texas at San Antonio, San Antonio, Texas 78249, USA}
\author{S. C. Burd}
\affiliation{Time and Frequency Division, National Institute of Standards and Technology, 325 Broadway, Boulder, Colorado 80305, USA}
\affiliation{Department of Physics, University of Colorado, Boulder, Colorado 80309, USA}
\author{D. H. Slichter}
\affiliation{Time and Frequency Division, National Institute of Standards and Technology, 325 Broadway, Boulder, Colorado 80305, USA}
\author{S. B. Libby}
\affiliation{Physics Division, Physical and Life Sciences, Lawrence Livermore National Laboratory, Livermore, California 94550, USA}
\author{D. Leibfried}
\affiliation{Time and Frequency Division, National Institute of Standards and Technology, 325 Broadway, Boulder, Colorado 80305, USA}
\date{\today}

\section{Transformation into classical frame of reference}
We are interested in an $n$-particle, 1-D  system with coordinates denoted $\hat{x}_{j}$ for position and $\hat{p}_{j}$ for momentum, with an uncoupled potential $V_{j}(\hat{x}_{j},t)$. This is represented with the Hamiltonian
\begin{equation}\label{eq:orig}
    \hat{H}_{l}(t) = \sum_{j}\frac{\hat{p}_{j}^{2}}{2m_{j}} + V_{j}(\hat{x}_{j},t).
\end{equation}
We assume that different modes $j$ are uncoupled and thus neglect this subscript from here on. We now apply a transformation into the classical frame of reference using two displacement operators,
\begin{eqnarray}
\hat{D}_{\hat{x}(t)} = \exp\Big\{\frac{-i}{\hbar}c(t)\hat{p}\Big\}
\end{eqnarray}
to transform into a frame centered around the classical position of the particle $c(t)$, and \begin{eqnarray}
\hat{D}_{\hat{p}}(t) = \exp\Big\{\frac{i}{\hbar}m\dot{c}(t)\hat{x} \Big\}
\end{eqnarray}
to transform into a frame centered around the classical momentum of the particle $m\dot{c}(t)$. Using the Baker-Campbell-Hausdorff theorem, we can combine these two transformations into one, giving
\begin{eqnarray}
\hat{D}(t) = \exp\Big\{\frac{i}{\hbar}\Big(m\dot{c}(t)\hat{x} - c(t)\hat{p} \Big) + i\frac{m\dot{c}(t)c(t)}{2\hbar}\Big\}.
\end{eqnarray}
We drop the third term in the above equation because it represents a global phase, yielding
\begin{eqnarray}
\hat{D}(t) = \exp\Big\{\frac{i}{\hbar}\Big(m\dot{c}(t)\hat{x} - c(t)\hat{p} \Big)\Big\}.
\end{eqnarray}
Noting that $\hat{D}^{\dagger}(t)\hat{x}\hat{D}(t) = \hat{x} + c(t)$ and $\hat{D}^{\dagger}(t)\hat{p}\hat{D}(t) = \hat{p} + m\dot{c}(t)$ \cite{wolf_1995}, we can transform into a frame of reference for $\hat{D}^{\dagger}(t)\ket{\psi(t)}=\ket{\phi(t)}$, where $\ket{\psi(t)}$ is the wave function in the original frame of reference given by Eq.~(\ref{eq:orig}). Up to a global phase, this gives 
\begin{eqnarray}
\hat{H}_{t} &=& \hat{D}^{\dagger}(t)\hat{H}_{l}(t)\hat{D}(t) + i\hbar\dot{\hat{D}}^{\dagger}(t)\hat{D}(t) \nonumber \\
&=& \frac{\hat{p}^{2}}{2m} + V(\hat{x}+c(t),t) + m\ddot{c}(t)\hat{x}.
\end{eqnarray}
Applying our assumption that $V(\hat{x}+c(t),t)$ is approximately quadratic around $\hat{x}=0$, and using Newton's equation of motion $m\ddot{c}(t) = -\partial V/\partial \hat{x}|_{\hat{x}=0}$, we obtain:
\begin{eqnarray}
\hat{H}_{t}(t) &=& \frac{\hat{p}^{2}}{2m} + \hat{x}\frac{\partial V}{\partial \hat{x}}|_{\hat{x}=0} + \hat{x}^{2}\frac{1}{2}\frac{\partial^{2}V}{\partial\hat{x}^{2}}|_{\hat{x}=0} - \hat{x}\frac{\partial V}{\partial \hat{x}}|_{\hat{x}=0} \nonumber \\
&=& \frac{\hat{p}^{2}}{2m} + \frac{1}{2}m\omega^{2}(t)\hat{x}^{2},
\end{eqnarray}
were we have substituted $\partial^{2}V/\partial\hat{x}^{2}|_{\hat{x}=0} = m\omega^{2}(t)$ in the second line.

\section{Second quantization based on initial well}
We wish to define the ladder operators for each problem based on the well the phonon mode initially describes, such that $\ket{\phi(0)}=\ket{0}$. The form of the Hamiltonian describing the dynamics of the $j^{th}$ mode is:
\begin{eqnarray}
    \hat{H}_{j}(t) &=& \frac{\hat{p}_{j}^{2}}{2m_{j}} + \frac{1}{2}m_{j}\Big[1 + \gamma_{j}(t) \Big]^{2}x_{j}^{2} \nonumber \\
    &=& \frac{\hat{p}_{j}^{2}}{2m_{j}} + \frac{1}{2}m_{j}\omega_{j}^{2}x_{j}^{2} + m_{j}\omega_{j}^{2}\gamma_{j}(t)\Big[1 + \frac{\gamma_{j}(t)}{2}\Big]x_{j}^{2}.
\end{eqnarray}
We can then substitute ladder operators based on the $\omega_{j}$ harmonic well:
\begin{eqnarray}
\hat{p}_{j} &=& i\sqrt{\frac{\hbar m_{j}\omega_{j}}{2}}\Big[\hat{a}^{\dagger}_{j} - \hat{a}_{j} \Big] \\
x_{j} &=& \sqrt{\frac{\hbar}{2m_{j}\omega_{j}}}\Big[\hat{a}_{j}^{\dagger} + \hat{a}_{j} \Big]\, ,
\end{eqnarray}
which gives
\begin{eqnarray}\label{eq:main_ham_supp}
\hat{H}_{j} &=& \hbar\omega_{j}\Big[\hat{a}^{\dagger}_{j}\hat{a}_{j} + \frac{1}{2}\Big] +  \frac{\hbar\omega_{j}\gamma_{j}(t)}{2}\Big[1 + \frac{\gamma_{j}(t)}{2} \Big]\Big[\hat{a}^{\dagger}_{j} + \hat{a}_{j} \Big]^{2}\, .
\end{eqnarray}

\section{Phonons in $\omega_{j}(t)\neq \omega_{j}(0)$ wells}

The mean number of phonons in a harmonic well defined by a frequency $\omega$ is
\begin{eqnarray}\label{eq:phonons_ch}
\braket{\hat{n}_{\omega}} = \frac{\braket{\hat{H}(t)}}{\hbar\omega} - \frac{1}{2}.
\end{eqnarray}
For the calculations in this work, we write our ladder operators $\hat{a}_{j}(\hat{a}_{j}^{\dagger})$ in terms of the initial well frequency $\omega_{j}(0)$ as $\omega_{j}(t) = [1 + \gamma_{j}(t)]\omega_{j}(0)$. Inserting this into the above equation gives
\begin{eqnarray}
\braket{\hat{n}_{\omega_{j}(t)}} = \Big[1 + \gamma_{j}(t)\Big]^{-1}\bra{\phi(t)}\Big[\Big(\hat{a}_{j}^{\dagger}\hat{a}_{j} + \frac{1}{2} \Big) + \frac{\gamma_{j}(t)}{2}\Big(1 + \frac{\gamma_{j}(t)}{2}\Big)\Big(\hat{a}^{\dagger}_{j} + \hat{a}_{j} \Big)^{2}\Big]\ket{\phi(t)} - \frac{1}{2}\, .
\end{eqnarray}

\section{Changing frequency of a well}
The squeeze operator acting on the $j^{th}$ mode of the system is defined as
\begin{eqnarray}\label{eq:change_freq_op}
\hat{U}_{c,j} = \exp\Big(\frac{r_{c,j}}{2}\Big[\hat{a}_{j}^{2 \dagger} - \hat{a}_{j}^{2} \Big]\Big),
\end{eqnarray}
where we have implicitly set $\phi_{c,j} = 0$ in the squeeze operator for simplicity. The transformation this operator induces on $\hat{a}_{j}$ is given by the Bogoliubov transformation \cite{Bogoliubov1958, agarwal_2012}
\begin{eqnarray}\label{eq:bogoliubov}
\hat{U}_{c,j}^{\dagger}\hat{a}_{j}\hat{U}_{c,j} = \hat{a}_{j}\cosh{r_{c,j}} + \hat{a}_{j}^{\dagger}\sinh{r_{c,j}}.
\end{eqnarray}
If we assume a starting state for the $j^{th}$ mode of $\ket{0}$, which is then acted upon by the Hamiltonian Eq.~(\ref{eq:main_ham_supp}), we take the expectation value as
\begin{eqnarray}\label{eq:changed_ham}
\braket{\hat{H}(t)} &=& \bra{0}\hat{U}_{c,j}^{\dagger}\Big(\hbar\omega_{j}\Big[\hat{a}^{\dagger}_{j}\hat{a}_{j} + \frac{1}{2}\Big] +  \frac{\hbar\omega_{j}\gamma_{j}(t)}{2}\Big[1 + \frac{\gamma_{j}(t)}{2} \Big]\Big[\hat{a}^{\dagger}_{j} + \hat{a}_{j} \Big]^{2} \Big)\hat{U}_{c,j}\ket{0} \nonumber \\
&=&\hbar\omega_{j}\bra{0}\hat{U}_{c,j}^{\dagger} \Big( f_{\gamma}\hat{a}_{j}^{\dagger}\hat{a}_{j} + \frac{1}{2}\Big[f_{\gamma}-1\Big]\Big[\hat{a}^{\dagger 2}_{j} + \hat{a}^{2}_{j} \Big] + \frac{1}{2}f_{\gamma}\Big)\hat{U}_{c,j}\ket{0},
\end{eqnarray}
where we have grouped operators. Here
\begin{eqnarray}
f_{\gamma} &\equiv & 1 + \gamma_{j}(t)\Big[1 + \frac{\gamma_{j}(t)}{2}\Big]. \nonumber \\
\end{eqnarray}
It is more straightforward to consider the Heisenberg equations of motions for the ladder operators, solving for the necessary $\hat{U}_{c,j}$ such that the above equation becomes
\begin{eqnarray}\label{eq:desired_heis_ham}
    \braket{\hat{H}(t)} &=& \bra{0}\Big(\hbar\omega_{j}\Big[1+\gamma_{j}(t)\Big]\Big[\hat{a}^{\dagger}\hat{a} + \frac{1}{2}\Big]\Big)\ket{0}  \nonumber \\
    &=& \frac{\hbar}{2}[1+\gamma_{j}(t)]\omega_{j}.
\end{eqnarray}
We can explicitly calculate the value of $r_{c,j}$ needed for this by taking our desired $\hat{H}(t)$ in the Heisenberg picture, Eq.~(\ref{eq:desired_heis_ham}), calculating $\hat{U}_{c,j}\hat{H}(t)\hat{U}_{c,j}^{\dagger}$, and solving for the value of $r_{c}$ such that this operation transforms the operator into that given by Eq.~(\ref{eq:changed_ham}); this is described by Eq.~(\ref{eq:bogoliubov}) with $r_{c,j}\rightarrow -r_{c,j}$. Doing this gives
\begin{eqnarray}
\hat{U}_{c,j}\hat{H}(t)\hat{U}_{c,j}^{\dagger} &=& \hbar\omega_{j}\Big[1+\gamma_{j}(t)\Big]\Big[(\hat{a}_{j}^{\dagger}\cosh r_{c,j} - \hat{a}_{j}\sinh r_{c,j})(\hat{a}_{j}\cosh r_{c,j} - \hat{a}^{\dagger}_{j}\sinh r_{c,j}) + \frac{1}{2}\Big] \nonumber \\
&=& \hbar\omega_{j}\Big[1 + \gamma_{j}(t) \Big]\Big(\hat{a}_{j}^{\dagger}\hat{a}_{j}\cosh 2r_{c,j} - \frac{1}{2}\Big[\hat{a}_{j}^{\dagger 2} + \hat{a}_{j}^{2} \Big]\sinh 2r_{c,j} + \frac{1}{2}\cosh 2r_{c,j} \Big),
\end{eqnarray}
which we match to Eq.~(\ref{eq:changed_ham}) and solve to obtain
\begin{eqnarray}\label{eq:r_value}
    r_{c,j} &=& -\frac{1}{2}\mathrm{arcsinh}\Big(\frac{\gamma_{j}(t)}{2}\Big[\frac{2+\gamma_{j}(t)}{1+\gamma_{j}(t)} \Big] \Big) \nonumber \\
    &=& -\frac{1}{2}\ln\Big\{1+\gamma(t_{f})\Big\}
\end{eqnarray}
It is easily shown that the terms $\propto \hat{a}_{j}^{\dagger}\hat{a}_{j}$ and $\propto \hat{I}$ match Eq.~(\ref{eq:changed_ham}) when the above equality is satisfied. While this derivation involves a lot of algebra, the end result is fairly intuitive when we consider that, in the Heisenberg picture, changing the basis of a harmonic oscillator from the eigenvectors of $\omega_{0}$ to $\omega(t_{f})$ means:
\begin{eqnarray}
\hat{x}_{j}&\rightarrow & \sqrt{\frac{\omega_{0}}{\omega(t_{f})}}\hat{x}_{j} =\{1 + \gamma(t_{f})\}^{-1/2}\hat{x}_{j} \nonumber \\
\hat{p}_{j}&\rightarrow & \sqrt{\frac{\omega(t_{f})}{\omega_{0}}}\hat{p}_{j} = \{1+\gamma(t_{f}) \}^{1/2}\hat{p}_{j}.
\end{eqnarray}
Combined with the fact that, under the Bogoliubov transformation above:
\begin{eqnarray}
\hat{U}_{c,j}^{\dagger}\hat{x}_{j}\hat{U}_{c,j} &\rightarrow& e^{r_{c,j}}\hat{x}_{j} \nonumber \\
\hat{U}_{c,j}^{\dagger}\hat{p}_{j}\hat{U}_{c,j} &\rightarrow& e^{-r_{c,j}}\hat{p}_{j},
\end{eqnarray}
gives rise to Eq.~(\ref{eq:r_value}).

\section{Parametric modulation}
The Hamiltonian for a harmonic oscillator whose frequency is undergoing sinusoidal modulation of amplitude $g$ is
\begin{eqnarray}
    \hat{H}(t) = \frac{\hat{p}^{2}}{2m} + \frac{1}{2}m\omega^{2}x^{2} + \hbar g\sin(2\omega t - \theta)\Big(\frac{x}{x_{0}}\Big)^{2},
\end{eqnarray}
where $x_{0}\equiv \sqrt{\hbar/(2\omega m)}$. Substituting ladder operators defined by the $\omega$ frequency well, and ignoring global phases, we get
\begin{equation}
    \hat{H}(t) = \hbar\omega\hat{a}^{\dagger}\hat{a} + \hbar g\sin(2\omega t - \theta)\Big(\hat{a}^{\dagger 2} + \hat{a}^{2} + 2\hat{a}^{\dagger}\hat{a} + 1 \Big).
\end{equation}
Transforming into the interaction picture with respect to $\hbar\omega\hat{a}^{\dagger}\hat{a}$ gives
\begin{eqnarray}
\hat{H}_{I}(t) &= & \frac{\hbar g}{2i}\Big(e^{i(2\omega t - \theta)} - e^{-i(2\omega t - \theta)} \Big)\Big(\hat{a}^{2 \dagger}e^{2i\omega t} + \hat{a}^{2}e^{-2i\omega t} + 2\hat{a}^{\dagger}\hat{a} + 1 \Big) \nonumber \\
&\simeq & \frac{i\hbar g}{2}\Big(\hat{a}^{\dagger 2}e^{i\theta t} - \hat{a}^{2}e^{-i\theta} \Big),
\end{eqnarray}
where we have made the rotating wave approximation in the second line.

\section*{Same-species ion separation Hamiltonian}
We write the Hamiltonian for two same-species ions in a Harmonic potential as
\begin{eqnarray}\label{eq:sep_orig_ham}
    \hat{H}_{l}(t) &=& \frac{\hat{p}_{1}^{2}}{2m} + \frac{\hat{p}_{2}^{2}}{2m} + \frac{1}{2}m\omega^{2}(t)\Big\{[\hat{x}_{1}-c_{f,1}(t)]^{2} + [\hat{x}_{2}-c_{f,2}(t)]^{2}\Big\} + \frac{ke^{2}}{\hat{x}_{2}-\hat{x}_{1}},
\end{eqnarray}
where $c_{f,j}(t)$ represents the minimum of the potential of the $j^{th}$ ion. Initially, these two terms represent the same well, centered at the origin of the lab-frame, $c_{f,j}(0) = 0$. After the ions are released, however, they represent the center of the two `catching' potentials for the ions, assumed to be far apart. We have also assumed that the time-dependence of the the confining strength that each ion sees is the same, represented here with $\omega(t)$. Transforming into the classical frame of reference with respect to each particle using the displacement operator $\hat{D}_{j}(t)$ and ignoring global phases gives
\begin{eqnarray}
\hat{H}_{t} \simeq \frac{\hat{p}_{1}^{2}}{2m} + \frac{\hat{p}^{2}_{2}}{2m} + \frac{1}{2}m\omega^{2}(t)[\hat{x}_{1}^{2}+ \hat{x}_{2}^{2}] + \frac{ke^{2}(\hat{x}_{2}-\hat{x}_{1})^{2}}{[c_{2}(t)-c_{1}(t)]^{3}},
\end{eqnarray}
where the frame transformation has removed all terms that are linear in position. In the above equation, we have Taylor-expanded the Coulomb term to $2^{nd}$-order, assuming that the value of $c_{2}(t)-c_{1}(t)$ is much larger than spatial extent of the motional wave functions of each ion at all times. We make the coordinate change
\begin{eqnarray}
 \hat{x}_{s} &\equiv &\frac{1}{2}(\hat{x}_{2}-\hat{x}_{1}) \nonumber \\ \hat{x}_{c} &\equiv & \frac{1}{2}(\hat{x}_{2}+\hat{x}_{1}) \nonumber \\
\hat{p}_{s} &\equiv & \hat{p}_{2}-\hat{p}_{1} \nonumber \\ \hat{p}_{c} &\equiv & \hat{p}_{2}+\hat{p}_{1},
\end{eqnarray}
which transforms Eq.~\ref{eq:sep_orig_ham} to
\begin{equation}
    \hat{H}(t) \simeq \frac{\hat{p}_{c}^{2}}{2M} + \frac{\hat{p}_{s}^{2}}{2M} + \frac{1}{2}M\omega^{2}(t)[\hat{x}_{c}^{2} + \hat{x}_{s}^{2}] + \frac{ke^{2}\hat{x}_{s}^{2}}{2c_{s}(t)^{3}},
\end{equation}
where $M\equiv 2m$. This represents an uncoupled differential equation with respect to the COM and STR modes, meaning we may divide the Hamiltonian into two parts, $\hat{H}_{s}(t)$ and $\hat{H}_{c}(t)$. Further, we can substitute the initial equilibrium configuration $c_{s}(0)=(ke^{2}/2M\omega_{0}^{2})^{-1/3}$. The resulting expressions for $\hat{H}_{s}(t)$ and $\hat{H}_{c}(t)$ are

\begin{eqnarray}
    \hat{H}_{c}(t) &=& \frac{\hat{p}_{c}^{2}}{2M} + \frac{1}{2}M\omega^{2}(t)\hat{x}_{c}^{2} \\
    \hat{H}_{s}(t) &=& \frac{\hat{p}_{s}^{2}}{2M} + \Big[\frac{1}{2}M\omega^{2}(t) + M\omega_{0}^{2}\Big\{\frac{c_{s}(0)}{c_{s}(t)}\Big\}^{3}\Big]\hat{x}_{s}^{2}.
\end{eqnarray}
Substituting $\omega(t) \equiv \omega_{0}[1+\gamma(t)]$, we find
\begin{eqnarray}
    \hat{H}_{c}(t) &=& \frac{\hat{p}_{c}^{2}}{2M} + \frac{1}{2}M\omega_{0}^{2}\Big\{1 +\gamma(t) \Big\}^{2} \hat{x}_{c}^{2} \\
    \hat{H}_{s}(t) &=& \frac{\hat{p}_{s}^{2}}{2M} + \frac{1}{2}M\omega_{0}^{2}\Big\{ [1+\gamma(t)]^{2}+ 2\Big\{\frac{c_{s}(0)}{c_{s}(t)}\Big\}^{3}\Big\}\hat{x}_{s}^{2} \nonumber \\
    &=& \frac{\hat{p}_{s}^{2}}{2M} + \frac{1}{2}M\omega_{0}^{2}\Big\{ 1 + \gamma_{c}(t)\Big\}^{2}\hat{x}_{s}^{2},
\end{eqnarray}
where in the last line, we have made the substitution
\begin{eqnarray}
\gamma_{c}(t) \equiv \Big\{[1+\gamma(t)]^{2} + 2\Big[\frac{c_{s}(0)}{c_{s}(t)} \Big]^{3} \Big\}^{1/2} - 1
\end{eqnarray}

\clearpage
\setcounter{equation}{0}
\renewcommand{\theequation}{S\arabic{equation}}

\section*{References}
\bibliography{biblio}